# The Registrar's Function in a Hybrid Society: AI Value Chain, Smart Data and the Concept of Property.

*Based on the panel "Imbuing Normative Principles on Digital Assets into Law" (FinteQC Conference, Québec, June 2026) — control · validity of public money · validity in the AI value chain.*

---

**Pompeu Casanovas (IIIA-CSIC) · Carmen Pastor Sempere (University of Alicante) · Marina Echebarría Sáenz (University of Valladolid)**

pompeu.casanovas@iiia.csic.es; carmen.pastor@ua.es; marina.echebarria@uva.es

**Abstract.** Artificial intelligence reaches the land registry not as another tool but as a value chain that turns data into intelligence and intelligence into economic value. This paper argues that the decisive legal move is to place validity — a functional, second-order concept — at the centre of that chain, letting rights, liability and supervision organise around it. It traces three impacts. Registry information becomes smart data, governed simultaneously by registry law, the GDPR, the European data acts and the AI Act. Control emerges as the operative concept for digital representations of real estate, whose proprietary effect depends on anchoring to the register. And, in a hybrid society of human and artificial agents, the registry becomes the public node of validity, with blockchain complementing rather than replacing it. Across three legal cultures, the registrar's value migrates from processing documents to guaranteeing validated data — making validity an asset for the Sustainable Development Goals.

## 1. Introduction: one structural question

Land registrars are used to being told that technology will change their work. Digitisation, electronic conveyancing, graphical databases and interoperable cadastres have been transforming registries for three decades. What is new about artificial intelligence is not that it adds another tool to that list. What is new is that AI arrives organised as a value chain — a sequence of activities that turns data into intelligence and intelligence into economic value — and that this chain now runs directly through the registry's three foundations: its data, the concept of property it administers, and its own institutional function.

The argument of this paper borrows a thesis developed for digital assets, public payments and AI regulation: in the new digital landscape, the decisive legal move is to place a functional, second-order concept at the centre of each field — control in private law, validity in public payments and in the AI value chain — and to let rights, liability and supervision organise around it.[1] Applied to land registration, the thesis becomes a reassurance and a warning at once. The reassurance: the registry already is, and has been for two centuries, an institution of validity — the public mechanism through which claims about land become legally certain. The warning: in a platform-driven economy fed by smart data, validity will be produced, contested and consumed in new ways, and the registry will keep its central position only if it occupies the centre of that chain deliberately.[2]

---

[1] C. Pastor Sempere, M. Echebarría Sáenz, and P. Casanovas, panel "Imbuing Normative Principles on Digital Assets into Law", FinteQC Conference, Québec (Canada), June 2026. On the value-chain frame generally, M. E. Porter, Competitive Advantage: Creating and Sustaining Superior Performance (Free Press, New York, 1985).

## 2. From the economic value chain to the AI value chain

The notion of a value chain comes from strategic management. Porter's generic chain (Figure 1) decomposes the firm into primary activities — inbound logistics, operations, outbound logistics, marketing and sales, service — sustained by support activities, with margin as the difference between the value created and its cost. Its quiet assumption matters here: in that picture, the ethical and legal dimension is a support activity — overhead at the edge of the chain, applied after value has been created. That is still how much of the technology industry books compliance.

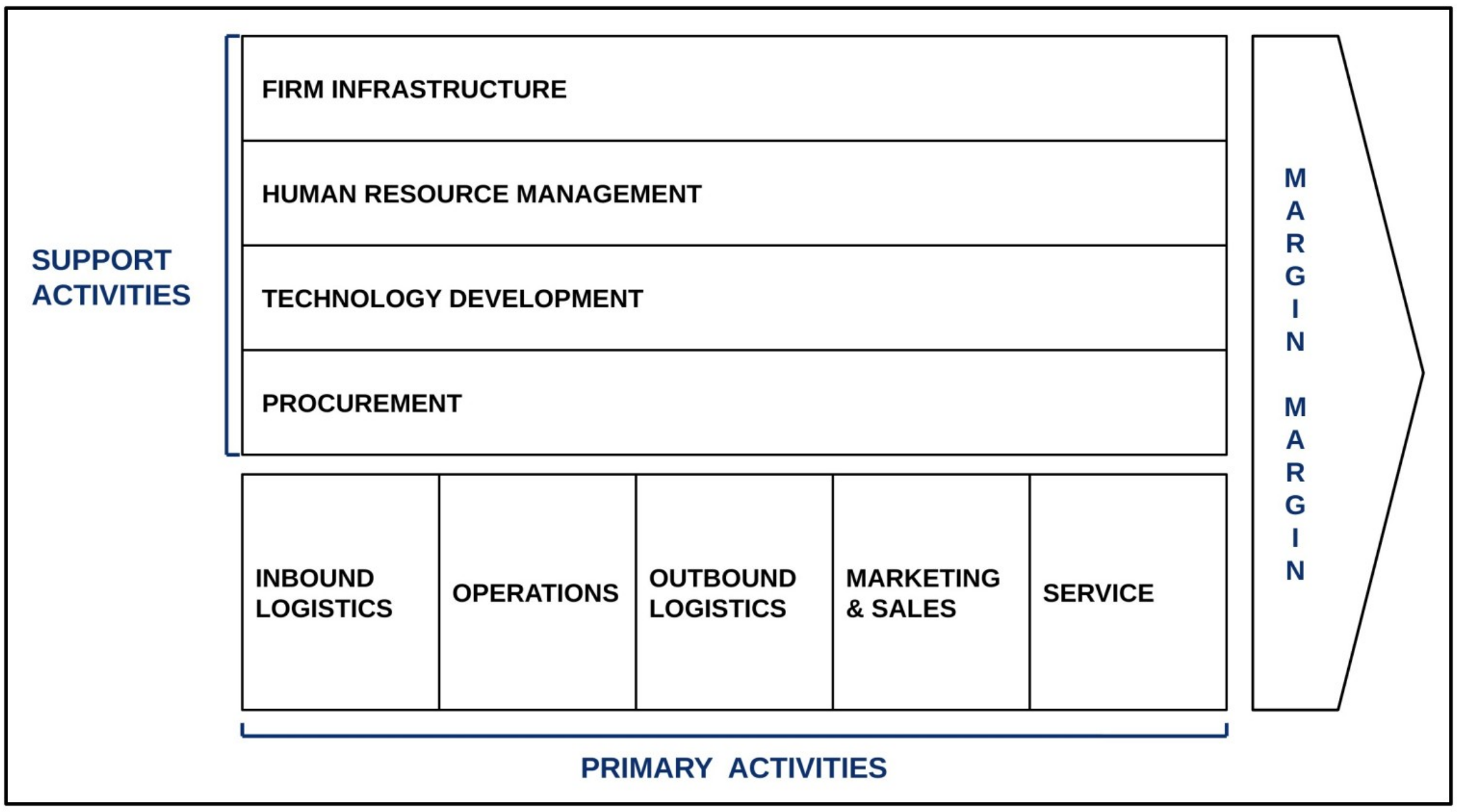

*Figure 1. Porter's generic value chain (1985, p. 37): the ethical-legal dimension as a mere support activity.*

The Delta Model of Hax and Wilde (Figure 2) corrects Porter in one decisive respect: value creation is understood through bonding, not through the product. The key idea is that competitiveness cannot be considered without including the customer: it is not merely a matter of increasing the firm's competitive advantage in the market, but of retaining the consumer or user of the service by generating value for them. Strategy moves along a triangle — best product, total customer solutions, system lock-in — and mature platform markets drift toward the third vertex. The land sector should read that drift literally: platform economics in property services, finance and AI tends structurally toward lock-in, and lock-in without a public anchor is exactly the MERS scenario examined below.[3]

[2] On the public-payments dimension of this thesis, see C. Pastor Sempere, "El ecosistema legal inteligente de pagos europeo: una aproximación metodológica y empírica", RDMV no. 38, 2026; on the AI value chain, P. Casanovas, "La cadena de valor ética y jurídica de la inteligencia artificial", ibid.

[3] A. C. Hax and D. L. Wilde II, The Delta Project: Discovering New Sources of Profitability in a Networked Economy (Palgrave Macmillan, New York, 2001); M. E. Porter, Competitive Advantage, cit. note 1.

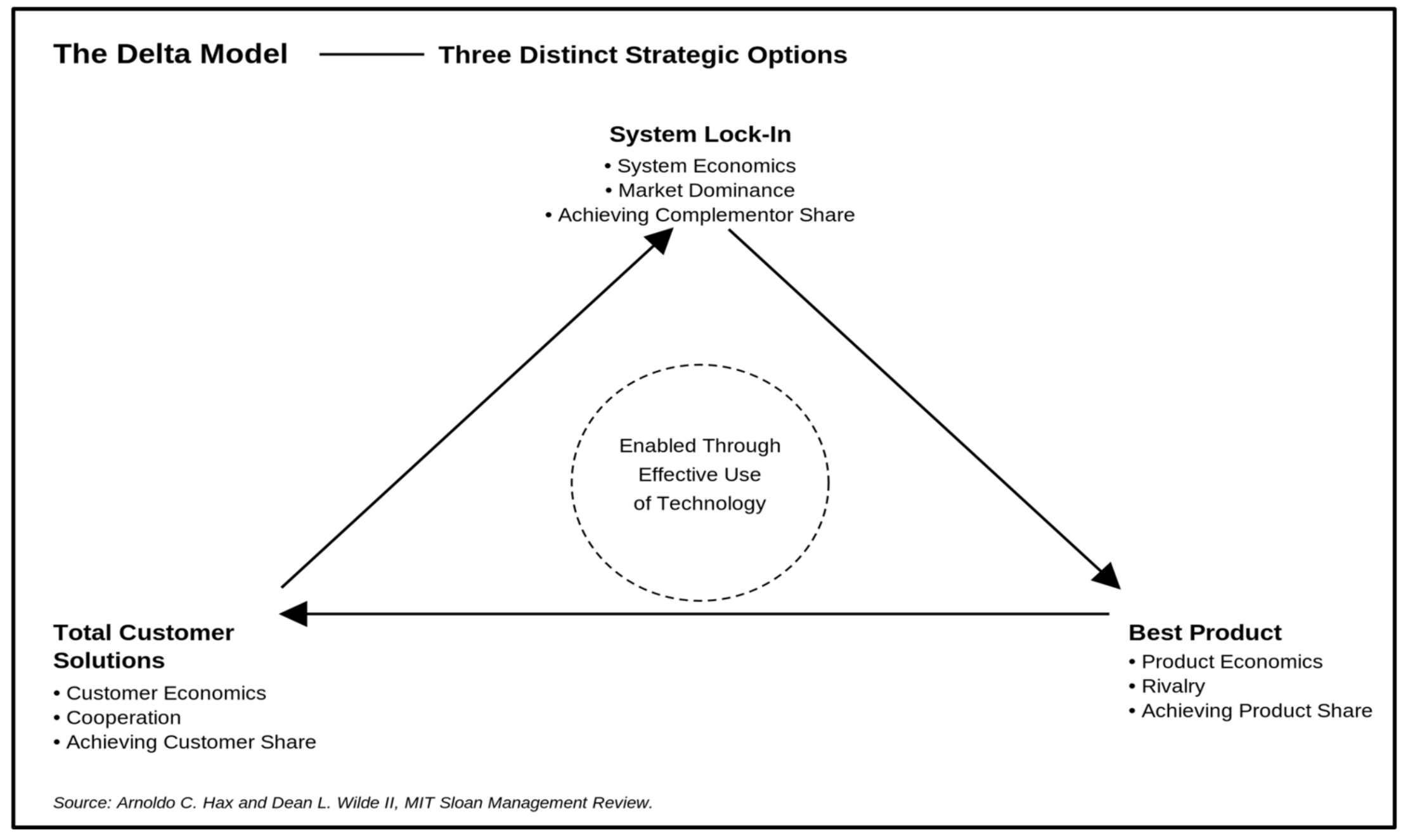

*Figure 2. The Delta Model triangle (Hax & Wilde, 2001): platform markets drift toward system lock-in.*

The AI value chain (AIVC) transposes this apparatus to intangibles that are not priced as products: a chain of activities running from data, through intelligence, to value, deployed across layered ecosystems — infrastructure, services, embedded uses — analogous to the four-layer stack of digital finance (Open Banking · BaaS · FinTech · embedded services).[4] Table 1 situates the AIVC among its neighbouring concepts: the supply value chain (goods and logistics), the global value chain (cross-border production) and the Delta model (service and bonding). What distinguishes the AIVC is its objective: not efficiency or lock-in, but validity, accountability and the trust it generates.[5]

*Table 1. Four related value-chain concepts.*

| Concept | Focus | Object | Objective |
|---|---|---|---|
| **SVC — supply value chain** | Goods & logistics | Physical products | Efficiency of supply |
| **GVC — global value chain** | Global lifecycle | Cross-border production | Jurisdictional coordination |
| **DMVC — Delta model** | Service logic | Customer relationship | Bonding & lock-in |
| **AIVC — AI value chain** | Data → intelligence → value | AI systems & their effects | Validity, accountability, trust |

[4] Cf. C. Pastor Sempere, "El ecosistema legal inteligente de pagos europeo: una aproximación metodológica y empírica", Revista de Derecho del Mercado de Valores no. 38, 2026.

[5] Cf. P. Casanovas Romeu, "La cadena de valor ética y jurídica de la inteligencia artificial: una propuesta operativa para la construcción del ecosistema financiero europeo", Revista de Derecho del Mercado de Valores no. 38, 2026.

The decisive move, then (Figure 3): the ethical and legal dimension is internalised as a primary activity of the chain itself, with legal validity at the centre — a second-order concept, formal and empirical at once, through which the first-order obligations scattered across the European AI Act, the GDPR, the data acts, eIDAS 2, MiCA and DORA acquire their meaning and their measure: compliance, liability, transparency, explainability, accountability, trust. The remainder of this paper applies that frame to the land registry.

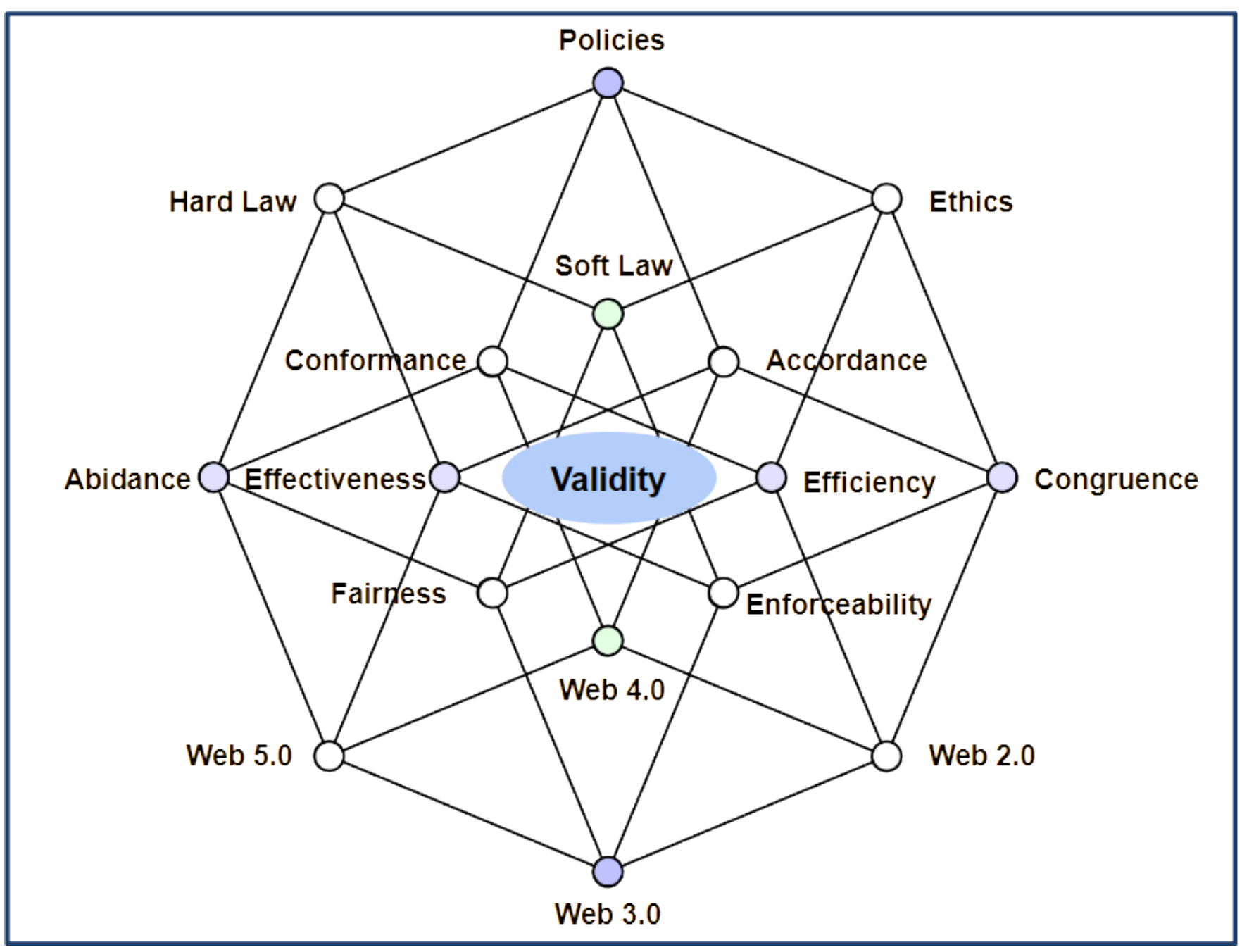

*Figure 3. Legal validity at the centre of the legal ecosystems. Source: Casanovas (2025).*

## 3. The impact on smart data

Begin with data, because that is where the AI value chain touches the registry first. A land register is not merely an archive of documents; it is a curated dataset of extraordinary quality — authenticated, structured, territorially complete and legally consequential. In the vocabulary of the AI value chain, registry data are the rare case of data that are already validity-bearing: every entry has passed a legal filter before becoming data at all.

AI changes both the supply side and the demand side of that dataset. On the supply side, machine learning is already industrialising the conversion of legacy records into structured, machine-readable form. HM Land Registry's Strategy 2025+ commits to automating straightforward transactions by 2030 and to near-instant, AI-assisted updates of the register by 2035, under human oversight; its data-science teams are using AI tools to migrate millions of pages of local land charges records, including handwritten and poorly scanned material.[6] In Spain, Law 11/2023 has made the registry fully electronic[7], and the Colegio de Registradores has opened an institutional debate on AI in the legal professions while releasing registry microdata for research.[8] Comparable programmes exist across the European Land Registry Association's membership.[9]

[6] HM Land Registry, Strategy 2025+: Digital services, expertise and accessible property information that unlock a better, faster and less stressful property market (November 2025), www.gov.uk/government/publications . The strategy commits to automating straightforward register changes by 2030, to near-instant, AI-assisted updates of the register under human oversight by 2035, and to accelerating the digitisation of local land charges records.

[7] Ley 11/2023, de 8 de mayo (BOE no. 110, 9 May 2023; BOE-A-2023-11022), whose Title IV makes the keeping of the Spanish Land, Commercial and Movable Property Registries fully electronic (new arts. 238 ff. Ley Hipotecaria), in force in this respect since 9 May 2024.

On the demand side, the platform economy consumes registry data at scale: lenders' automated valuation models, insurers' risk engines, short-term rental platforms, energy and infrastructure planners, and — increasingly — AI agents that search, summarise and act on titles. "Smart data" names the result of this double movement: registry information that is machine-readable, interoperable, linked to geospatial and cadastral layers, and consumed by algorithms rather than read by humans.

Two legal consequences follow. First, quality and provenance become legal questions, not merely technical ones. When an AI system makes a credit, planning or purchase decision on registry data, the registry's accuracy guarantee silently extends into the algorithmic decision; under the EU AI Act, several of these downstream uses (creditworthiness assessment, essential services) are classified high-risk, with data-governance duties that reach back to the data source.[10] Second, the European data acquis — the GDPR, the Data Governance Act, the Data Act, the open-data regime for high-value datasets — pulls registry data toward openness and reuse, while the registry's own logic pulls toward controlled, fee-based, purpose-bound publicity.[11] Managing that tension — open enough to feed the chain, controlled enough to remain authoritative and privacy-compliant — is the first new task of the registrar.

## 4. The impact on the legal concept of property and real estate

The second impact is conceptual. Private-law scholarship on digital assets has converged on a pivot: control, not ownership, is the analytical key to assets that live in information systems — the exclusive power to use, exclude others and transfer these abilities to third parties, which travels across legal systems where the inherited categories of property and possession strain.[12] Land seems immune: it is the paradigmatic corporeal thing. But real estate increasingly circulates through digital representations — tokenised shares of property companies, fractional investment platforms, smart-contract escrows, digital twins of buildings — and at the level of the representation, the control logic applies in full. Who controls the token, the account, the key? Which register, at which moment, determines standing and good faith?[13]

[8] Colegio de Registradores de España (CORPME), conference "La Inteligencia Artificial y las instituciones jurídicas" (2024), revistaregistradores.es. Registry microdata are available to accredited researchers through the Banco de España's BELab data laboratory under the CORPME–Banco de España agreement (Resolución of 26 July 2019, BOE-A-2019-11791), and selected open datasets through opendata.registradores.org.

[9] European Land Registry Association (ELRA), www.elra.eu .

[10] Regulation (EU) 2024/1689 of 13 June 2024 laying down harmonised rules on artificial intelligence (AI Act), OJ L, 12.7.2024: see Annex III, point 5(b) (creditworthiness assessment) and point 5(a) (essential services), and Art. 10 (data and data governance).

[11] Regulation (EU) 2016/679 (General Data Protection Regulation); Regulation (EU) 2022/868 (Data Governance Act); Regulation (EU) 2023/2854 (Data Act); Directive (EU) 2019/1024 on open data and the re-use of public sector information, and Commission Implementing Regulation (EU) 2023/138 on high-value datasets.

[12] Cf. M. Echebarría Sáenz, Section I, panel "Imbuing Normative Principles on Digital Assets into Law", FinteQC Conference, Québec (Canada), June 2026. Cf. also UNIDROIT, Principles on Digital Assets and Private Law (Rome, 2023), esp. Principle 6 (control), www.unidroit.org; Law Commission of England and Wales, Digital Assets: Final Report (Law Com No 412, 2023). See Echebarría Sáenz, M. "Principle 6: Control", in UNIDROIT Principles on Digital Assets and Private Law: Context, Content, Perspectives, Cham, Switzerland, 2025, pp. 313–334, open access at https://link.springer.com/book/10.1007/978-3-032-11001-5#toc

[13] On the EU framework for the representation of value and rights by crypto-assets, Regulation (EU) 2023/1114 on markets in crypto-assets (MiCA); for tokenised financial instruments, Regulation (EU) 2022/858 (DLT Pilot Regime).

The three great legal cultures absorb this pressure differently, and registrars should see the differences clearly, because they define three distinct futures for the profession.

In the civil-law tradition (Spain, Germany, the Latin notarial-registral systems), registration is the product of an ex-ante legal review — the Spanish calificación registral[14] — performed by an independent legal professional, and the register enjoys strong presumptions of accuracy and public faith. Property is a closed system of rights in rem (numerus clausus); a token over land is not itself a right in rem, but at most a representation of, or a claim to, one. The civil-law route to tokenisation therefore runs through the registry: the token acquires real-property effects only if and when the register says so. AI here is an instrument inside the qualification — triage, anomaly detection, document comparison — while the decision on validity remains a legal judgment.

In the common-law registration systems outside the United States (England and Wales, the Torrens jurisdictions of Australia, Canada and beyond), title derives from registration itself, backed by a state guarantee and an indemnity fund.[15] These systems are institutionally closest to full automation: if title is whatever the register says, an AI-maintained register is conceivable in principle, and HM Land Registry's 2035 ambition points that way. The legal-cultural question is what remains of judgment when alteration, rectification and indemnity — the points where the guarantee is tested — confront machine-made entries. Common law adapts by accretion: the Law Commission's recognition of a third category of personal property for digital objects, defined in substance by a test of control, shows the method.[16]

The United States is the outlier: there is, in general, no title registration but a recording of deeds at county level, with priority rules (race, notice, race-notice) and no state guarantee; certainty is supplied privately, by title search and title insurance. AI lands there on the private side: automated title search, instant underwriting, remote online notarisation, and tokenisation experiments that effectively create private, platform-held registers. The American lesson for everyone else is double. Positively, it shows how fast a data-driven market can move. Negatively, the experience of MERS — the private mortgage-registration system whose opacity complicated the foreclosure crisis — shows what happens when the record of rights migrates into a private platform without a public validity anchor.[17] Each registered mortgage received a unique 18-digit Mortgage Identification Number (MIN) that stayed with the loan throughout its life, however many times it changed hands. During the 2008–2012 crisis, although the system was designed to cut costs, MERS was found listed in the public records as the "nominal beneficiary" of millions of mortgages, obscuring who actually owned the loan; this generated extensive litigation, and the legal basis on which it operated proved poorly founded.[18]

---

See "Governance and Control of Data and Digital Economy in the European Single Market. Legal framework for new digital assets, identities and data spaces" in Cham, Switzerland, Springer, 2024, https://doi.org/10.1007/978-3-031-74889-9_13#DOI

[14] Art. 18 Ley Hipotecaria (consolidated text approved by the Decree of 8 February 1946).

[15] Land Registration Act 2002 (England and Wales). The Torrens model originates in the Real Property Act 1858 (South Australia).

[16] Law Commission, Digital Assets: Final Report (Law Com No 412, 2023), recommendation 1 and chs. 3–4; Digital Assets as Personal Property: Supplemental Report and Draft Bill (Law Com No 416, 2024), implemented through the Property (Digital Assets etc) Bill introduced in Parliament in September 2024.

[17] Cf. C. L. Peterson, "Foreclosure, Subprime Mortgage Lending, and the Mortgage Electronic Registration System", 78 University of Cincinnati Law Review 1359 (2010); also C. L. Peterson, "Two Faces: Demystifying the Mortgage Electronic Registration System's Land Title Theory", 53 William & Mary Law Review 111 (2011).

[18] See J. Puzzanghera, "Mortgage database's murky legal status adds another wrinkle to foreclosure mess", Los Angeles Times, 21 October 2010 (reporting that courts in several states held MERS — named on some 20 million home loans

This pivot is no longer confined to scholarship and soft law; it is being enacted. In England and Wales, the Property (Digital Assets etc) Act 2025 (Royal Assent, 2 December 2025) confirms that a thing is not prevented from being the object of personal property rights merely because it is neither a thing in possession nor a thing in action, recognising a third category of personal property. The Dubai International Financial Centre has gone further: its Digital Assets Law (DIFC Law No. 2 of 2024, in force 8 March 2024) is the first statute to set out comprehensively the legal characteristics of a digital asset as property — classifying it as intangible property, neither a thing in possession nor a thing in action (Article 9), and making control the operative criterion, since, absent contrary evidence, control implies superior legal title (Article 11), in line with Principle 6 of the UNIDROIT Principles on Digital Assets and Private Law; it also introduces self-executing "Coded Contracts" into DIFC contract law. The convergence of the Law Commission, UNIDROIT, the 2025 UK Act and the DIFC statute confirms the direction of travel: across legal cultures, control is becoming the operative concept for assets that live in information systems, and registration systems supply the validity anchor that turns control over a token into an effect on the right itself.

## 5. The registry in a hybrid society: role and function

This brings us to the third impact: the institution itself. A hybrid society is one in which humans and artificial agents act together — where mortgage offers, valuations, due diligence and soon conveyancing steps are performed by machines under human supervision. In such a society, the scarce resource is not information; platforms have plenty. The scarce resource is validity: the assurance that a representation of rights corresponds to legally effective reality and will be upheld erga omnes.

Seen through the AI value chain, the registry is therefore not a back-office to be automated away; it is the public node of validity in the chain — the institutional point where data about land stop being claims and become law. Three functional shifts follow.

First, from documents to data: the object of registration practice progressively becomes the validated data stream — structured deeds, qualified electronic signatures and seals (eIDAS 2 and the European Digital Identity Wallet)[19], interoperable cadastral geometry — rather than the paper instrument that once carried it. This shift has been described with precision from within the Colegio de Registradores itself: eIDAS 2 entails moving from the mere functional equivalence between the physical and electronic worlds — the logic of the electronic document — to the digital wallet conceived as a container of verifiable data linked to their authentic source.[20] For data on land and on companies, that authentic source is the Register: the wallet does not create validity but carries credentials whose authenticity the Register guarantees — drastically reducing the costs of documentary verification in sectors such as banking and insurance and reinforcing — far as mortgagee of record for lenders and investors — not to be the true owner, questioning its standing to foreclose).

[19] Regulation (EU) 2024/1183 of 11 April 2024 amending Regulation (EU) No 910/2014 as regards establishing the European Digital Identity Framework ("eIDAS 2"), OJ L, 30.4.2024.

[20] D. Calvo González-Vallinas, "Del documento electrónico a la cartera digital", *elEconomista*, June 2026. The author — a land registrar and Director of European Affairs of the Colegio de Registradores de España — characterises eIDAS 2 as the evolution from the functional equivalence between the physical and electronic document towards the digital wallet conceived as a container of verifiable data linked to their authentic source, with the EUDI Wallet (European Digital Identity Wallet) at the centre of the system.

from eroding it — the Register's position as the public node of validity. The same logic has been projected onto the commercial sphere: because the legality control that the commercial registrar exercises when granting access to information ensures that the registered data are reliable and verified at source, the Commercial Register operates as an authentic source within the meaning of eIDAS 2 — so that its data may be incorporated into the digital wallet and circulate as verifiable credentials — and registral publicity confers on them the effectiveness and the erga omnes opposability that a merely private credential lacks. This legality control, far from being an analogue encumbrance, is reinforced and extended at the European level: Directive (EU) 2025/25 on the digitalisation of company law imposes a preventive legality control prior to registration and projects it across the entire life-cycle of the company, confirming that dual notarial-and-registral control systems are compatible with Union law and requiring merely declaratory registers to adapt to that standard.[21]

Second, from *ex-post* correction to *ex-ante* design: just as public subsidy money builds its lawfulness into the conditions of issuance rather than policing it afterwards[22], registries will increasingly encode the conditions of valid entry into the systems themselves — legal governance by design — with AI checking completeness, chains of title and inconsistencies before a human registrar decides the hard cases. AI handles volume; the registrar's qualification concentrates on judgment: capacity, representation, fraud, conflicting rights, the protection of the weaker party.

Third, from gatekeeper to guarantor: when entries are prepared by machines, the registrar's signature changes meaning — it certifies not only the result but the validity of the process, including the human oversight, traceability and explainability that the AI Act demands of high-risk systems.[23] Liability, indemnity and rectification doctrines will have to be re-mapped onto a chain in which providers, deployers and the registry share the production of an entry.

The comparative cultures again diverge in emphasis: civil-law systems will defend the qualification as the human core and use AI around it; title-registration systems will push automation furthest and stress the guarantee; the United States will let the market automate and discover, perhaps painfully, the price of lacking a public anchor. But the direction is common: the registrar's value migrates from processing to validating.

A live, mature example of this arrangement already operates in Australia. Property Exchange Australia (PEXA), created in 2010 to deliver the Council of Australian Governments' national e-conveyancing solution, is a privately operated platform — an Electronic Lodgment Network Operator supervised under the Electronic Conveyancing National Law and the model rules of the Australian Registrars' National Electronic Conveyancing Council (ARNECC) — that has become the country's dominant means of settling and lodging property dealings.[24] Its objective is a single

[21] D. Calvo González-Vallinas, op. cit. (note 20), develops this thesis in the commercial sphere: Commercial Registers, as authentic sources, supply data that circulate as verifiable credentials under Regulation (EU) 2024/1183 (eIDAS 2), reinforced by the opposability inherent in registral publicity. On preventive legality control, see Directive (EU) 2025/25 of the European Parliament and of the Council of 19 December 2024 amending Directives 2009/102/EC and (EU) 2017/1132 as regards further expanding and upgrading the use of digital tools and processes in company law (OJ L, 10 January 2025); its transposition must be completed by 31 July 2027 at the latest.

[22] C. Pastor Sempere, "Public Embedded Finance" and the ALCOIN proof of concept (Bons Digitals JoveMercat, Ajuntament d'Alcoi, 2025), Part II of the FinteQC panel cited in note 1.

[23] Regulation (EU) 2024/1689 (AI Act), Arts. 13 (transparency) and 14 (human oversight) for high-risk AI systems.

[24] Property Exchange Australia (PEXA), www.pexa.com.au . Formed in 2010 (originally National e-Conveyancing Development Ltd) to deliver the Council of Australian Governments' national e-conveyancing initiative, PEXA operates as an Electronic Lodgment Network Operator under the Electronic Conveyancing National Law, within the Model Oper-

national channel through which lawyers, conveyancers and financial institutions prepare, sign and lodge instruments and settle the funds in near-real time. Its decisive feature for the present argument is infrastructure integration: PEXA is the official digital bridge to government, plugged directly into the State Revenue Offices for stamp duty and into the state Land Titles Offices for registering changes of ownership. Yet the boundary holds — the platform performs the processing, while the public Land Titles Office remains the point at which an entry becomes legally effective: PEXA is the channel, the register is the validity.

## 6. Blockchain, smart legal ecosystems and the boundary of the registrable

What, then, of blockchain — the technology most often presented as the registry's replacement?[25] Its function must be stated precisely. A distributed ledger produces a specific kind of certainty: integrity and chronology by consensus — entries that are time-stamped, traceable, tamper-resistant and, through smart contracts, programmable. That is functionally close to what registries have always done at the surface: ordering claims in time and protecting the record against manipulation. But the resemblance hides a categorical difference. Blockchain guarantees that the record has not changed; it cannot guarantee that what was recorded was ever valid — that the seller owned, that the power of attorney existed, that the transaction was lawful. An invalid act recorded on-chain is not cured but perpetuated. The registry's contribution lies exactly in the layer blockchain lacks: the ex-ante judgment of legality that turns a claim into a right. The mature conclusion of a decade of pilots — Sweden's Lantmäteriet, Georgia's National Agency of Public Registry, HM Land Registry's Digital Street[26] — is therefore complementarity, not substitution: blockchain as an evidence, audit and transport layer; the registry as the validity layer. EU law has begun to institutionalise this division of labour: eIDAS 2 now regulates electronic ledgers as trust services, and the European Blockchain Services Infrastructure (EBSI) is built to carry public-sector attestations, not to replace the authorities that issue them.[27]

More precisely, eIDAS 2 (Regulation (EU) 2024/1183) introduces electronic ledgers as a qualified trust service (Section 11): under Article 45k an electronic ledger may not be denied legal effect or admissibility as evidence merely because it is in electronic form, and the data records of a qualified electronic ledger enjoy a presumption of unique and accurate sequential chronological ordering and of integrity, with the conditions for qualified status set out in Article 45l. This is the European legal vehicle for the anchoring proposed below: it lends the ledger's integrity a rebuttable legal presumption, while leaving the validity of the underlying right to the register. A concrete implementation of this trust-service model is the BlockchainFUE network — a Valencian cooperative society, the first in Spain to operate a public blockchain, whose founding members ating Requirements and Model Participation Rules administered by the Australian Registrars' National Electronic Conveyancing Council (ARNECC).

[25] Cf. Oleksii Konashevych, Tokenization of Real Estate on Blockchain, Doctoral Dissertation, Alma Mater Università di Bologna and Universitat Autònoma de Barcelona, Erasmus Mundus Joint International Doctoral Degree in Law, Science and Technology, 2020.

[26] Lantmäteriet (Sweden), blockchain testbed with ChromaWay, Telia and Kairos Future (2016–2018): Kairos Future, The Land Registry in the Blockchain — Testbed (March 2017); National Agency of Public Registry of Georgia, blockchain land-titling project with the Bitfury Group (from 2016); HM Land Registry, "Digital Street" research and development programme (2018–2019), hmlandregistry.blog.gov.uk.

[27] Regulation (EU) 2024/1183 (eIDAS 2), introducing electronic ledgers as trust services (new arts. 45k–45l of Regulation (EU) No 910/2014); European Commission, European Blockchain Services Infrastructure (EBSI), digital-strategy.ec.europa.eu.

include the Colegio de Registradores de la Propiedad, Mercantiles y de Bienes Muebles de España (which holds the secretaryship of its governing board) — examined as an eIDAS 2 qualified electronic-ledger trust service by Alamillo-Domingo, and the network on which the AL-COIN proof of concept was run.[28] That the national registrars are themselves cooperative members of the ledger is a concrete instance of registrars taking part in the governance of a supervised ecosystem from the design stage.

Blockchain rarely arrives alone. It comes embedded in smart legal ecosystems: combinations of smart contracts, tokens, digital identity wallets, oracles and, increasingly, AI agents, in which part of the applicable rules is executed by code — compliance and legal governance by design.[29] These ecosystems will interrelate with the role, functions and practice of registrars in at least three ways.

Upstream, as an input channel: transactions will reach the registry as structured, signed, machine-verifiable data — smart deeds presented through eIDAS wallets — so the qualification will operate on code and data as much as on prose, and registrars will need the technical literacy to read both. At this point, the European Digital Identity Wallet (EUDI Wallet) operates as the container through which those verifiable credentials reach the qualification: the registrar moves from collating documents to verifying electronic attributes against their source, with the consequent saving in verification costs.

Downstream, as an execution environment that treats the registry as its oracle of validity: a smart contract that releases the price upon confirmation of registration; a mortgage discharged automatically upon verified payment; a token whose transferability is switched on or off by the state of the folio.

And laterally, as supervised ecosystems: permissioned networks for conveyancing or public payments in whose governance registrars, notaries and administrations participate from the design stage — the same ex-ante logic that public embedded finance applies to subsidy money.

[28] I. Alamillo-Domingo, "La red BlockchainFUE como servicio de confianza de libro mayor electrónico conforme al Reglamento eIDAS 2", RDMV no. 38, 2026.

[29] P. Casanovas Romeu, "The Ethical and Legal AI Value Chain", Part III of the FinteQC panel cited in note 1; P. Casanovas, L. de Koker and M. Hashmi, "Law, Socio-Legal Governance, the Internet of Things, and Industry 4.0: A Middle-Out/Inside-Out Approach", J — Multidisciplinary Scientific Journal 5(1) (2022), 64–91. Also, for smart manufacturing: P. Casanovas Romeu, "Building a Smart Legal Ecosystem for Industry 5.0", W. Barfield, Y. Hsuan, U. Pagallo, The Cambridge Handbook of the Law, Policy, and Regulation for Human–Robot Interaction , Cambridge University Press, 2014, pp. 145 – 170.

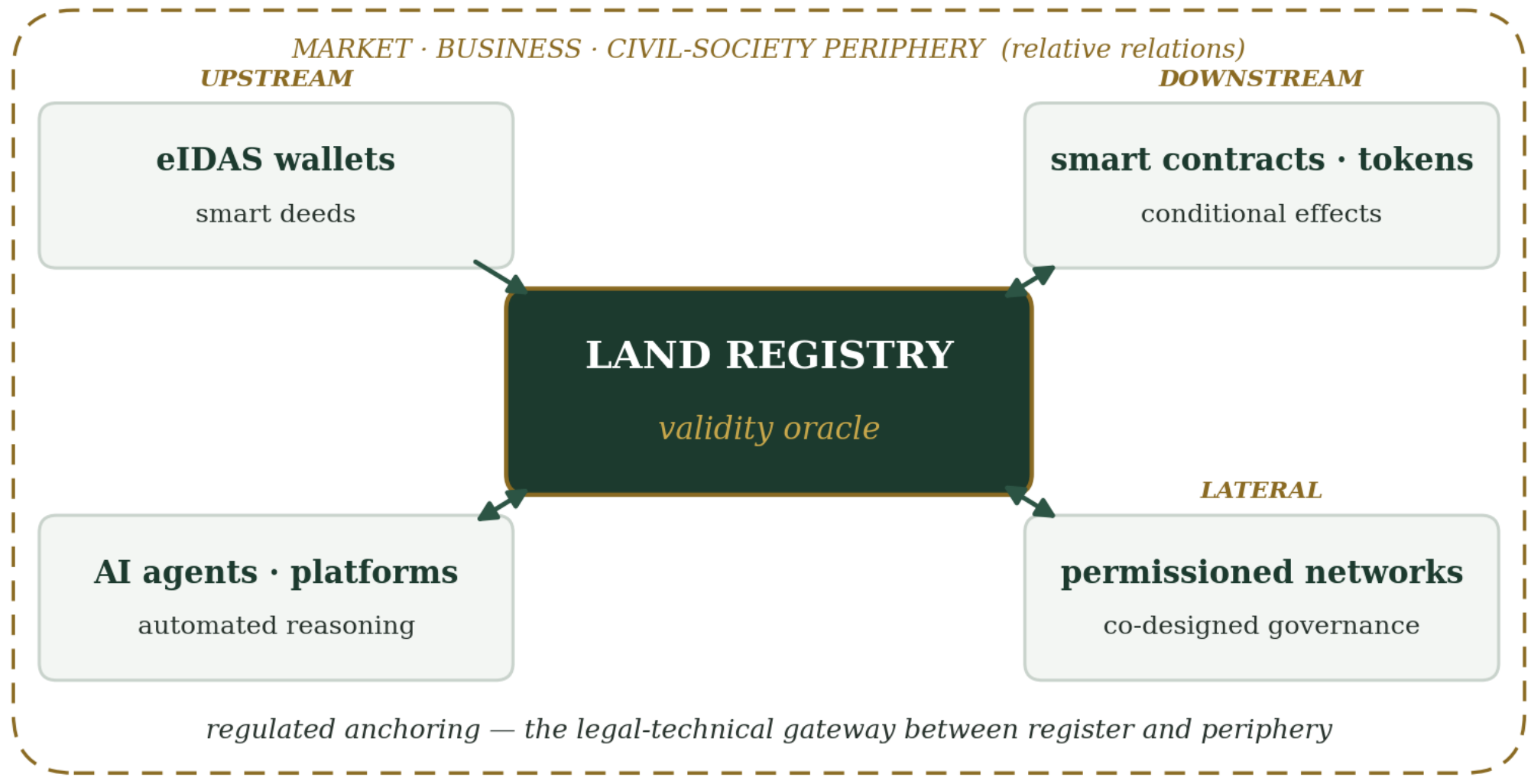

*Figure 4. The smart legal ecosystem: the land registry as oracle of validity, with regulated anchoring to the private periphery.*

The hard question is the boundary: what should be registered, and what should be left to the market, business and civil-society relationships? The criterion is not technological but functional, and it is the classic one: erga omnes effect. What belongs in the registry is what must bind everyone — ownership and the numerus clausus of rights in rem, priorities, charges and prohibitions, and the anchoring points of digital representations (which token, if any, counts as connected to which folio). What can remain outside is the dense web of relative relations that platforms and contracts manage well enough among the parties: fractional investment schemes, revenue-sharing and use tokens, short-term rental arrangements, DAO governance, reputational and ESG attributes of buildings. Two symmetrical mistakes mark the edges of this boundary. One is privatisation by drift: letting platform ledgers harden into de facto registers of rights without a public anchor — the MERS lesson once more. The other is over-registration: dragging every digital relation into the folio, which would overload the institution, freeze innovation and dissolve the selectivity that makes registry publicity meaningful. To which we should add another important factor: the increased cost of intermediation in the market due to the effect of the tariff.

The sound architecture is a flexible, authoritative public core of validity surrounded by a thick private periphery, interoperating through standards — with anchoring, the legal-technical act that connects a token or smart contract to an entry, as the regulated gateway between the two.

Seen this way, smart legal ecosystems do not bypass the registrar; they presuppose one. Code can execute rules, but it cannot decide what is lawful; markets can price claims, but they cannot make them opposable to third parties. In a landscape of interoperating ledgers, the registrar's qualification becomes the bridge between code and law — and the register itself, the reference ledger that the other ledgers trust.

A live example now exists. In March 2025 the Dubai Land Department (DLD) launched the pilot phase of its Real Estate Tokenisation Project under the "REES" initiative — the first time in the region that a land-registration authority has tokenised property title deeds — in collaboration with the Virtual Assets Regulatory Authority (VARA), the Dubai Future Foundation and the Central Bank of the UAE. The tokens are minted on a public ledger, and the tokenisation infrastructure is integrated directly with the DLD so that the on-chain record and the official register remain synchronised: the register stays the legal "ground truth", and each token must map to the DLD's cadastral identifiers (plot and title-deed numbers). This is precisely the architecture argued for here — the register as the oracle of validity, the token deriving its real-property effect from its anchoring to the official entry. Phase II, opened in February 2026, activated secondary-market resale of the tokens, with the DLD projecting a tokenised market of some AED 60 billion, around 7% of Dubai's property transactions, by 2033. The model also illustrates the boundary of the registrable: the proprietary effect flows from the registry anchor, while the trading and fractional-ownership dimension sits in the regulated private periphery, under VARA's asset-referenced virtual asset (ARVA) framework. Such tokenised real estate falls largely outside the harmonised perimeter of MiCA (Regulation (EU) 2023/1114), which addresses crypto-assets, asset-referenced and e-money tokens rather than the property-law status of immovables — confirming that the real-property effect of a token remains, for now, a matter of national property and registry law connected through the validity anchor.

## 7. AI, land registries and the SDGs

This frame answers the workshop's question directly, because validity is an SDG asset. On the social side, tenure security and access to credit rest on validated, opposable rights: registry data lower information asymmetry for lenders and protect the weaker party, and AI-driven accessibility — plain-language interfaces, assistance for vulnerable groups, bridging the digital divide — is only trustworthy when it operates on validated data (SDGs 1, 8, 10 and 16). On the environmental side, fully electronic, paper-free registries reduce the sector's footprint, while energy-efficiency and emissions attributes of buildings linked to the folio, and geospatial AI over cadastral data, make the register a carrier of verified green information for protected areas and sustainable cities (SDGs 7, 11, 13 and 15).[30]

Institutionally, validity is trust in institutions made operational (SDGs 16 and 17): transparent, explainable, human-overseen AI inside the registry, and a public anchor that keeps platform lock-in accountable. The SDG contribution of land administration in the age of AI is therefore not a by-product of digitisation; it is the direct effect of keeping a public validity layer at the centre of the chain.

## 8. Conclusions

1. AI reaches the registry as a value chain — data → intelligence → value — not as a single tool; the registry's strategic position is its place in that chain, not any particular technology.
2. Registry data are becoming smart data: machine-readable, interoperable and consumed by algorithms. Their quality, provenance and governance are now legal questions, gov-erned simultaneously by registry law, the GDPR, the European data acts and the AI Act's high-risk regime.

[30] UNECE Working Party on Land Administration (WPLA) and Colegio de Registradores de España, workshop "How can AI and the digitalization in the land sector support achieving the SDGs?", Decanato de Cataluña del Colegio de Registradores, Barcelona, 9–10 June 2026 (programme v. 4 June 2026).

3. For digital representations of real estate, control — the exclusive power to use, exclude and transfer — is the operative concept; tokens acquire real-property effect only through connection to a validity anchor, which in registration systems is the register itself.
4. Blockchain complements rather than replaces the registry: it guarantees the integrity of the record, not the validity of the right. Smart legal ecosystems — with smart contracts, tokens, wallets, AI agents — presuppose a public validity layer and will use the registry as their authoritative oracle; the boundary of the registrable remains functional: erga-omnes effects belong in the register, while relative, contractual relations are left to market, business and civil society, connected through regulated anchoring.
5. The three legal cultures define three trajectories: civil law integrates AI inside *ex-ante* qualification; common-law title registration automates the register under a state guarantee; the United States automates privately through title search and insurance, with the MERS experience as a standing warning.
6. In a hybrid, platform-driven society the scarce good is validity, not information; the registry's function is to be the public node of validity in the AI value chain.
7. The registrar's role shifts from processing documents to guaranteeing and explaining validated data: AI absorbs volume; human judgment concentrates on qualification, oversight and the protection of third parties and weaker parties.
8. Validity is an SDG asset: tenure security and access to credit (social), a paper-free sector and verified green attributes of buildings (environmental), and trustworthy, accountable institutions (institutional) all follow from keeping a public validity layer at the centre of the AI value chain.
9. None of this is self-executing: it requires investment in data quality, machine-readable standards, eIDAS-based identity, AI-governance frameworks inside registries — and a doctrinal effort to re-map liability and indemnity onto human–machine co-produced entries.

The standards layer is itself becoming an object of public strategy. The United Kingdom's "Shaping Tomorrow: The UK's Digital Standards Strategy (2026–2030)" (DSIT, 17 June 2026) treats digital standards as strategic infrastructure that complements regulation and underpins interoperability — a reminder that the machine-readable standards on which the register's interoperability depends are increasingly shaped in standards-development bodies, and that public authorities, registrars among them, have an interest in helping to write them.

## 9. A question for discussion

Australia already demonstrates how far the private market can go: PEXA, a privately operated platform, now settles and lodges the overwhelming majority of the country's property dealings — more than twenty thousand settlements a week, on infrastructure plugged directly into the state Land Titles Offices and now designated as critical national infrastructure — yet the public register still remains the point at which an entry becomes legally effective. The platform took the channel; the register kept the validity. If, by 2035, an AI system can in addition update that register in near-real time, and a PEXA-scale platform can wrap the whole transaction in a faster, cheaper private substitute, the line between channel and validity will be tested as never before: what, then, is the irreducible core of the registrar's qualification — the legal judgment that neither the machine may make nor the market may buy — and how should we write it into law before the technology, or the market, writes it for us?

***Acknowledgments:*** *Proyecto Prometeo LegalCripto (CIPROM/2022/26, Generalitat Valenciana; Proyecto Civic AI (CAI), i-link 2026/27 CSIC; Unidad Asociada CSIC-UAB. We thank Pilar Verdejo and Teresa Touriñán for the invitation to participate in the UNECE WPLA & Registradores de España Workshop, as well as the registrars for their feedback. We also thank Patrick Keyzer and Louis de Koker for their valuable comments. CLAUDE has been used for layout and content editing. Responsibility for this content rests with the authors.*

## References

### European Union legislation

### Spain

### UNITED KINGDOM AND COMMONWEALTH

### UNITED ARAB EMIRATES

### INTERNATIONAL INSTRUMENTS

### SCHOLARSHIP AND OTHER SOURCES